# A High-resolution Clock Phase Shifter Circuitry for ALTIROC

X. Huang,[a] C. de La Taille,[b] D. Gong,[a,1] C. Liu,[a] T. Liu,[a] M. Morenas,[b] N. Seguin-Moreau,[b] J. Ye,[a] and L. Zhang[a]

[a] *Southern Methodist University,*
*Dallas, TX 75275, USA*

[b] *OMEGA/CNRS/Ecole Polytechnique,*
*91128 PALAISEAU CEDEX, France*
*E-mail:* dgong@smu.edu

ABSTRACT: A high-resolution clock phase shifter is implemented to adjust the phase of multiple clocks at 40 MHz, 80 MHz, or 640 MHz in the ALTIROC chip. The phase shifter has a coarse-phase shifter and a fine-phase shifter to achieve a step size of 97.7 ps and an adjustable range of 25 ns. The fine delay unit is based on a Delay Locked Loop (DLL) operating at 640 MHz. The phase shifter is fabricated in a 130 nm CMOS process. The area of the phase shifter is 725 µm × 248 µm. The Differential Non-Linearity (DNL) and the Integral Non-Linearity (INL) are ±0.6 LSB and ±0.75 LSB, respectively. The jitter from -25 ºC to 20 ºC is less than 15.5 ps (RMS), including the contributions from the FPGA clock source and the PLL. The power consumption is 11.2 mW.

---

[1] Corresponding author.

## Contents

## 1. Introduction

The High Granularity Timing Detector (HGTD) will accurately measure the time of the tracks with a time resolution better than 50 ps in RMS per hit to mitigate the pile-up effects after the ATLAS Phase-II upgrade [1]. The HGTD is based on a Low Gain Avalanche Detector (LGAD) technology with excellent timing performance [2]. The HGTD uses a dedicated ASIC, ATLAS LGAD Timing Integrated Read-Out Chip (ALTIROC), to read out the LGAD. ALTIROC will be bump-bonded to the LGAD sensor. The full size of ALTIROC has 15 × 15 channels, corresponding to the 225 pads of the LGAD sensor. ALTIROC measures the Time Of Arrival (TOA) with respect to the 40 MHz bunch-crossing clock and the Time Over the Threshold (TOT) of the analog signal as an estimation of the signal amplitude to correct for the time-walk effect offline.

To ensure that the readout channel works with the right timing in the small timing window, we must precisely adjust the phases of the clock signals. Both Versions 1 [3] and 2 [4] of ALTIROC have internal Phase-Locked Loops (PLLs) and clock phase shifters afterward. The block diagram of the clock generator (PLL + phase shifter) in ALTIROC2 is shown in **Figure 1**. The PLL produces a 40 MHz clock (PLL_40MHz) and a 640 MHz clock (PLL_640MHz) based on the reference 40 MHz clock recovered from lpGBT. The phase shifter provides five phase independently adjustable clocks, two at 40 MHz, one at 80 MHz, and the other two at 640 MHz. The minimum step size of each phase-adjustable clock is 1/16 of the 640 MHz clock period or 97.7 ps in a 25 ns range.

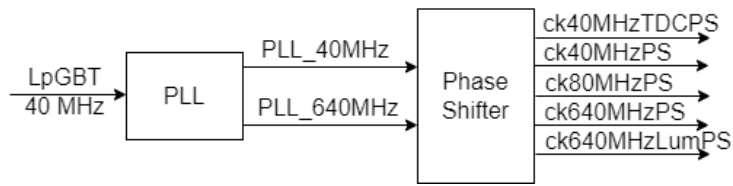

**Figure 1.** Block diagram of clock generator in ALTIROC2.

## 2. Design

The block diagram of the phase shifter is shown in **Figure 2**. The phase shifter is implemented with a coarse-delay unit, a fine-delay unit, and a re-sampler. The coarse delay unit consists of a differentiator and two 4-bit counters. The coarse delay unit is driven by the clocks PLL_40MHz and PLL_640MHz from the PLL. The differentiator is composed of two D flip flops (DFF1 and DFF2) and an AND (AND1) gate. The differentiator generates a pulse signal with a period of 25 ns to reload periodically and simultaneously the coarse phase settings to the two counters. The periodic reloading of the two counters eliminates the accumulation of Single Event Upset effects. One counter produces a 40 MHz clock (ck40MHz1). The other counter generates a 40 MHz clock (ck40MHz2) and an 80 MHz clock (ck80MHz). The output clocks of the two counters are synchronized and phase-adjustable with a step size of 1.5625 ns. The fine delay unit is based on a Delay Locked Loop (DLL) [5] [6] operating at 640 MHz. The DLL consists of a Voltage-Controlled Delay Line (VCDL), a phase detector, a charge pump, and a first-order Low Pass Filter (LPF). The VCDL outputs 16 tapped clocks. The phase difference between adjacent tapped clocks is 97.7 ps. Two 16-to-1 multiplexers (MUX16) share one VCDL to select the desired phases for 640 MHz clocks. The re-sampler is used to align the three coarse-phase clocks with one of the fine-phase shifted clocks (ck640MHz1) to obtain a step size of 97.7 ps.

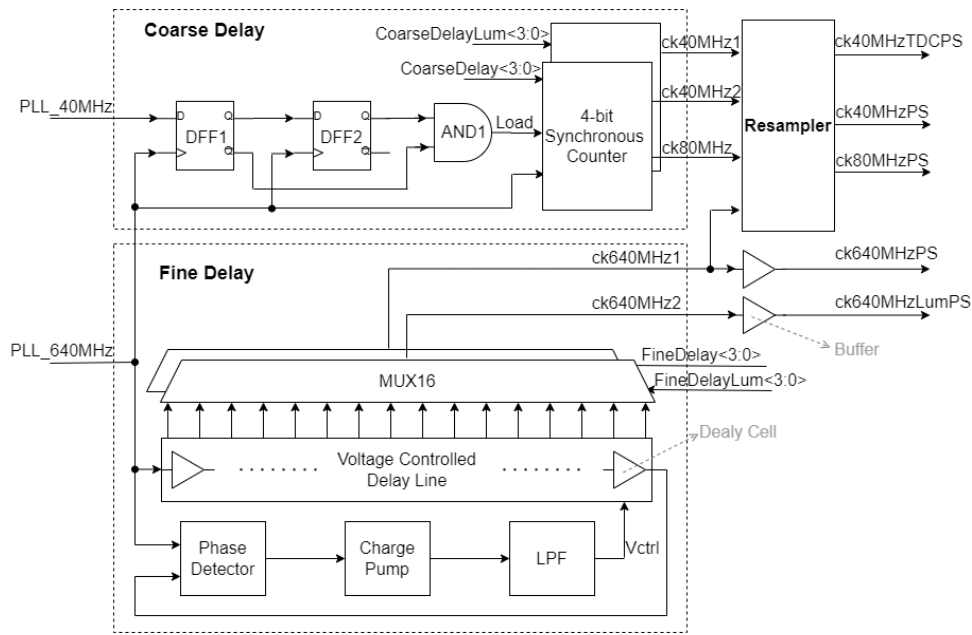

**Figure 2.** Block diagram of the phase shifter in ALTIROC2.

The VCDL consists of 17 identical delay cells and divides 1.5625 ns into 16 equaling phases. Each delay cell is composed of two identical half delay cells. The schematic of the half delay cell is shown at the left of **Figure 3**. A PMOSFET PM0 and a current source I0 are a simplified bias circuit to provide a reference voltage for the half delay cell. Two PMOSFETs PM1 and PM2 and two NMOSFETs NM1 and NM2 form a current starving inverter. The delay of the half delay cell is controlled by a control voltage (Vctrl). The direct output (OUT1) of the half delay cell goes to the next half delay cell. The buffered output (OUT2) is used in the phase detector. The gated output (OUT3) is connected to the phase-selection multiplexer.

The phase detector compares the rising edges of the buffered outputs out of the first delay cell and the 17[th] delay cell of the VCDL. The phase detector produces two pulse signals according to the phase difference and drives the charge pump to adjust a control voltage Vctrl. The control voltage Vctrl ensures that the interval of the VCDL input and output is a period of the 640 MHz clock. The LPF stabilizes Vctrl. The 16-to-1 multiplexer (MUX16) selects the gated clocks with a constant phase difference from the VCDL. The basic cell of MUX16 is a 2-to-1 multiplexer. The schematic of the 2-to-1 multiplexer is shown at the right of **Figure 3**. The multiplexer cell is symmetrical. For the left branch, PM1, PM2, NM1, and NM2 are used to select the input IN0, and PM2 and NM1 perform as switches to control the output.

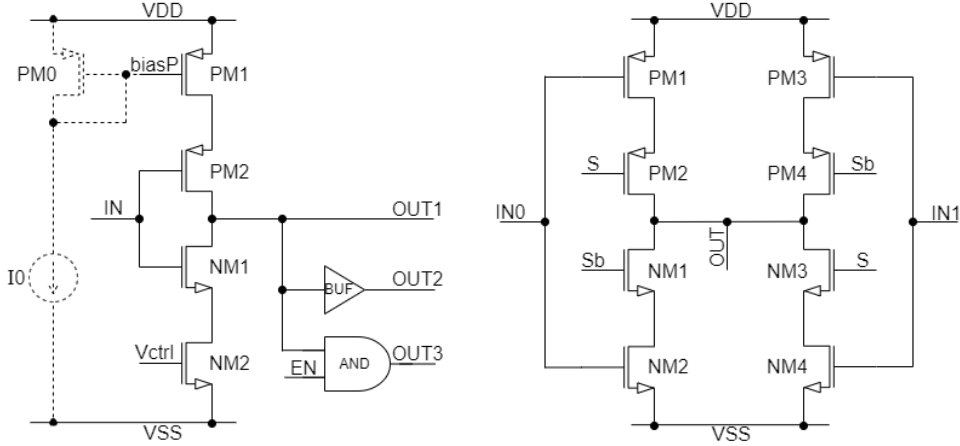

**Figure 3.** Schematics of the half delay cell and the 2-to-1 multiplexer.

During the resampling process, potential metastability occurs resulting from the edges of ck640MHz1 and coarse-phase clocks (ck40MHz1, ck40MHz2, and ck80MHz). A re-sampler is used to relieve the metastability effect. The schematic of the re-sampler is shown in **Figure 4**. The re-sampler consists of three duplicated resampling circuits. Take the re-sampler of the coarse-phase clock of 80MHz (ck80MHz) as an example. An inverter (INV) and a DFF1 produce a pre-processed coarse-phase clock (ck80MHz_pre) by sampling the ck80MHz clock at the falling edge of ck640MHz1. A small phase difference exists between the ck80MHz clock and the ck80MHz_pre clock. The ck80MHz clock and the ck80MHz_pre clock are selected by a control signal (ckSel). The potential metastability can be found by scanning the 16 phases of the clock 640MHz1. The default output of the MUX2 is the ck80MHz clock. The ck80MHz_pre clock is selected when metastability is predicted.

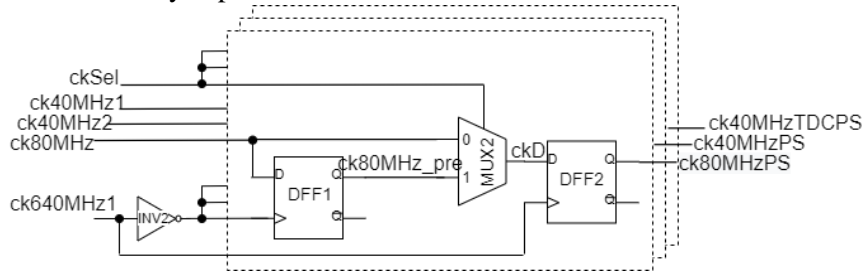

**Figure 4.** Schematic of the re-sampler.

The phase shifter is fabricated in a 130 nm CMOS process. The area of the phase shifter is 725 μm × 248 μm. A screenshot of the layout is shown in **Figure 5**.

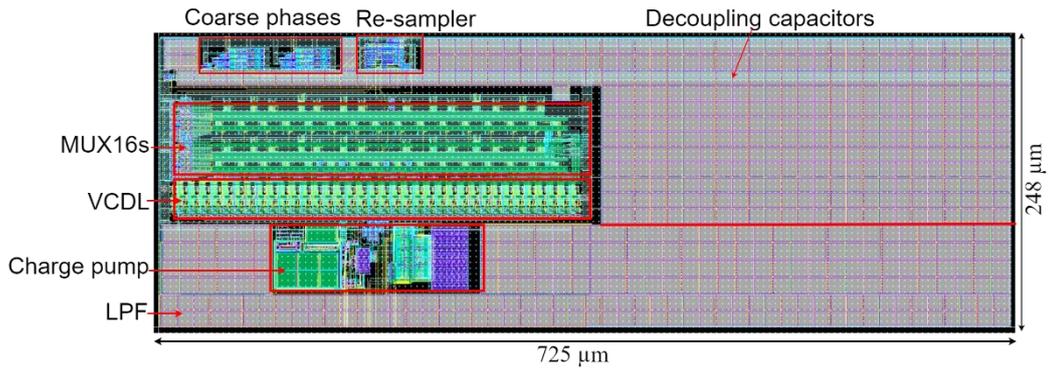

**Figure 5.** Screenshot of the phase shifter layout.

## 3. Test

The phase shifter was tested along with the PLL in ALTIROC. The block diagram and a photograph of the test setup are shown in **Figure 6**. An FPGA (AMD Xilinx, Zynq-7000 SoC ZC706 Evaluation Kit, part No. EK-Z7-ZC706-G) provided a 40 MHz clock and communicated with the ALTIROC2 chip. The internal PLL generated a 40 MHz and a 640 MHz clock for the phase shifter. The output clocks of the phase shifter were tested by using a high-speed oscilloscope (LeCroy Waverunner 8254).

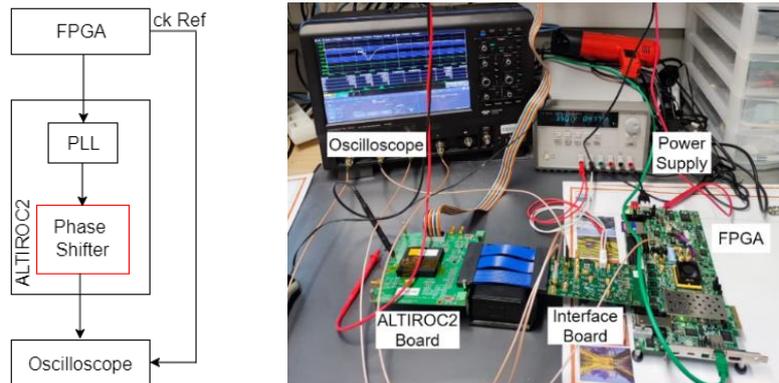

**Figure 6.** Block diagram and photograph of the test setup.

The nonlinearity and the jitter performance of the phase shifter were tested. The nonlinearity is shown in **Figure 7**. The Differential Non-Linearity (DNL) and the Integral Non-Linearity (INL) are ±0.6 LSB and ±0.75 LSB, respectively. The clock jitter for all 256 phases is measured at four temperatures of -25 ℃, -10 ℃, 10 ℃, and 20 ℃. The histogram of the measured jitter is shown in **Figure 8**. The jitter from -25 ℃ to 20 ℃ is less than 15.5 ps (RMS), including the contributions from the FPGA clock source and the PLL. The power consumption of the phase shifter is 11.2 mW.

The irradiation test of the phase shifter was carried out along with the ALTOROC2 chip in an X-ray irradiator (Precision X-ray irradiation, X-RAD iR160 biological irradiator) at CERN. The ALTIROC2 test board was irradiated to the targeted Total Ionizing Dose (TID) of 2 MGy at a dose rate of 29.9±0.5 kGy/h at 22 ℃. The phase shifter functioned normally with unchanged phase noise measurements before and after the irradiation.

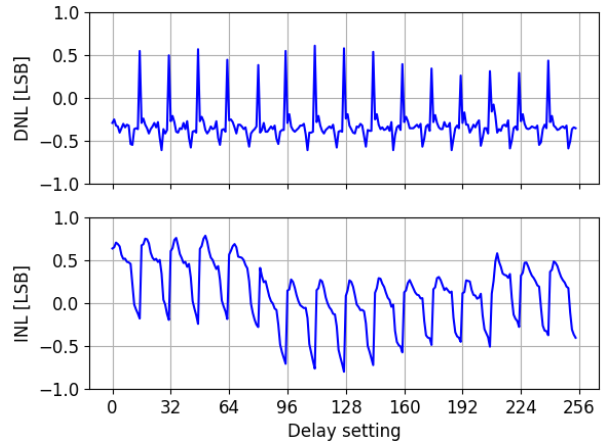

**Figure 7.** Nonlinearity of the phase shifter.

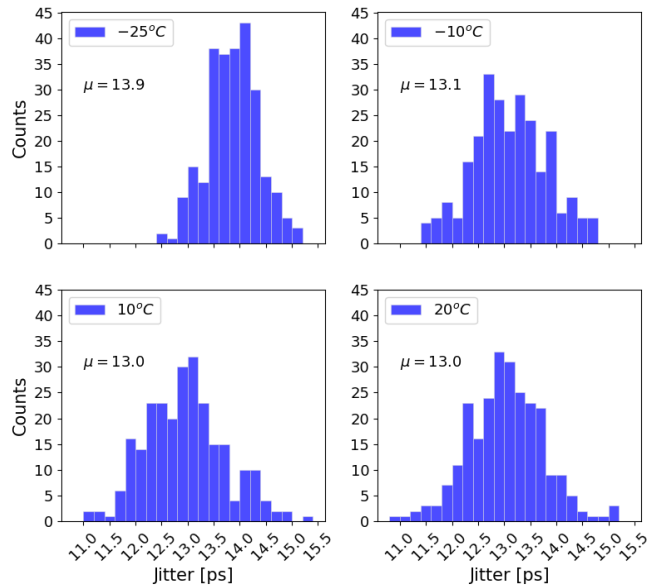

**Figure 8.** Jitter performance measurements at different temperatures.

## 4. Conclusion

A high-resolution clock phase shifter is implemented to adjust the phase of multiple clocks at 40 MHz, 80 MHz, or 640 MHz in the ALTIROC chip. The phase shifter has a coarse-phase shifter and a fine-phase shifter to achieve a step size of 97.7 ps and an adjustable range of 25 ns. The fine delay unit is based on a DLL operating at 640 MHz. The phase shifter is fabricated in a 130 nm CMOS process. The area of the phase shifter is 725 µm × 248 µm. The Differential Non-Linearity (DNL) and the Integral Non-Linearity (INL) are ±0.6 LSB and ±0.75 LSB, respectively. The jitter at four temperatures is less than 15.5 ps (RMS), including the contributions from the FPGA clock source and the PLL. The power consumption is 11.2 mW.

## Acknowledgments

This work is supported by the NSF and the DOE Office of Science, SMU's Dedman Dean's Research Council Grant.

# References

[1] M.P. Casado, *A High-Granularity Timing Detector for the ATLAS Phase-II upgrade.* Nuclear Instruments and Methods in Physics Research Section A: Accelerators, Spectrometers, Detectors and Associated Equipment, 1 Jun 2022, 1032:166628.

[2] S.M. Mazza, *A High-Granularity Timing Detector (HGTD) for the Phase-II upgrade of the ATLAS detector.* 2019 JINST 14 C10028.

[3] N. Seguin-Moreau et al., *ALTIROC, a 25-ps time resolution ASIC for ATLAS HGTD*, talk given at TWEPP 2019 Topical Workshop on Electronics for Particle Physics, Santiago de Compostela, Spain, 2 – 6 September 2019.

[4] M. Morenas et al., *Performance of ALTIROC2 readout ASIC with LGADs for ATLAS HGTD picosecond MIP timing detector,* talk given at TWEPP 2022 Topical Workshop on Electronics for Particle Physics, Bergen, Norway, 19 – 23 September 2022.

[5] D. Yang et al., *A High-Resolution Clock Phase-Shifter in a 65ánm CMOS Technology*, Proceedings of International Conference on Technology and Instrumentation in Particle Physics (TIPP 2017), Springer Proceedings in Physics, vol 212, Springer, Singapore.

[6] G. Jovanovic et al., *Delay locked loop with linear delay element*, TELSIKS 2005-2005 uth International Conference on Telecommunication in Modern Satellite, Cable and Broadcasting Services 2005 Sep 28 (Vol. 2, pp. 397-400).